\documentclass[graybox]{svmult}

\usepackage{mathptmx}       
\usepackage{helvet}         
\usepackage{courier}        
\usepackage{type1cm}        
\usepackage{makeidx}         
\usepackage{graphicx}        
\usepackage{multicol}        
\usepackage[bottom]{footmisc}

\usepackage{amsmath,amssymb}

\DeclareMathOperator{\res}{res}
\DeclareMathOperator{\const}{const}

\newcommand{\set}[2]{\ensuremath{\left\{{#1}\ \middle|\ {#2}\right\}}}

\makeindex             

\begin{document}

\title*{Generalization of Risch's Algorithm to Special~Functions}
\author{Clemens G. Raab}
\institute{Clemens G. Raab \at Deutsches Elektronen-Synchrotron, DESY, Platanenallee 6, 15738 Zeuthen, Germany, \email{clemens.raab@desy.de}}
%
%
\maketitle

\abstract{Symbolic integration deals with the evaluation of integrals in closed form. We present an overview of Risch's algorithm including recent developments. The algorithms discussed are suited for both indefinite and definite integration. They can also be used to compute linear relations among integrals and to find identities for special functions given by parameter integrals. The aim of this presentation\footnote{based on a lecture at the summer school on quantum field theory in July 2012 at RISC, Hagenberg} is twofold: to introduce the reader to some basic ideas of differential algebra in the context of integration and to raise awareness in the physics community of computer algebra algorithms for indefinite and definite integration.}

\section{Introduction}
\label{RaabSec:intro}

In earlier times large tables of integrals were compiled by hand \cite{RaabRef:Integraltafel2,RaabRef:GR,RaabRef:MagnusOberhettinger,RaabRef:PrudnikovEtAl}. Nowadays, computer algebra tools play an important role in the evaluation of definite integrals and we will mention some approaches below. Tables of integrals are even used in modern software as well. Algorithms for symbolic integration in general proceed in three steps. First, in computer algebra the functions typically are modeled by algebraic structures. Then, the computations are done in the algebraic framework and, finally, the result needs to be interpreted in terms of functions again. Some considerations concerning the first step, i.e., algebraic representation of functions, will be part of Section~\ref{RaabSec:AlgRep}. A brief overview of some approaches and corresponding algorithms will be given below. We will focus entirely on the approach using differential fields. Other introductory texts on this subject include \cite{RaabRef:NormanInt,RaabRef:BronsteinTut}. Manuel Bronstein's book on symbolic integration \cite{RaabRef:Bronstein} is a standard reference. The interested reader is referred to one of these for more information. A recent version of Risch's algorithm will be presented in Section~\ref{RaabSec:Alg}. The subtle issues of the last step, i.e., translating the algebraic result to a valid statement in the world of functions, will not be dealt with here.

\subsection{Parametric Integration}
\index{parametric integration}

Integration of functions can be done in two variants: indefinite and definite integration, which are closely related via the fundamental theorem of calculus. On the one hand, an indefinite integrals still is a function in the variable of integration and is nothing else than the antiderivative of a given function $f(x)$. On the other hand, a definite integral is the value
\[
 \int_a^bf(x)\,dx
\]
resulting from integrating the function $f(x)$ over the given interval $(a,b)$. Another difference between the two is that in general it is easy to verify an indefinite integral just by differentiating it, whereas in general it is hard to verify the result of a definite integral without recomputing it.

For the evaluation of definite integrals many tools may be applied to transform them to simpler integrals which are known or can be evaluated easily: change of variable, series expansion of the integrand, integral transforms, etc. As mentioned above by the fundamental theorem of calculus it is obvious that we can use indefinite integrals for the evaluation of definite integrals. It is well known that for a function $g(x)$ with $g^\prime(x)=f(x)$ we have
\[
 \int_a^bf(x)\,dx=g(b)-g(a).
\]
This fact has also been exploited in order to evaluate definite integrals for which a corresponding indefinite integral is not available in nice form. We give an overview of this method, which will be the main focus for computing definite integrals in the present paper. If the integral depends on a parameter, we can differentiate the parameter integral with respect to this parameter and obtain an integral that might be evaluated more easily. Under suitable assumptions on the integrand we have
\[
 \frac{d}{dy}\int_a^bf(x,y)\,dx=\int_a^b\frac{df}{dy}(x,y)\,dx,
\]
which is called \emph{differentiating under the integral sign}. A related paradigm, known as \emph{creative telescoping}, is used in symbolic summation to compute recurrences for parameter dependent sums, see \cite{RaabRef:ZeilbergerCT} for instance. Based on these two principles Almkvist and Zeilberger \cite{RaabRef:AlmkvistZeilberger} were the first to propose a completely systematic way for treating parameter integrals by differentiating under the integral sign by giving an algorithm to compute differential equations for parameter integrals with holonomic integrands. They gave a fast variant of it for hyperexponential integrands, which may also be used for computing recurrences for such parameter integrals. From a very general point of view the underlying principle might be understood as combination of the fundamental theorem of calculus and the linearity of the integral in the following way. If for integrable functions $f_0(x),\dots,f_m(x)$ and constants $c_0,\dots,c_m$ the function $g(x)$ is an antiderivative such that
\[
 c_0f_0(x)+\dots+c_mf_m(x)=g^\prime(x),
\]
then we can deduce the relation
\[
 c_0\int_a^bf_0(x)\,dx+\dots+c_m\int_a^bf_m(x)\,dx=g(b)-g(a)
\]
among the definite integrals $\int_a^bf_i(x)\,dx$ provided they exist. Both the functions $f_i(x)$ and the constants $c_i$ may depend on additional parameters, which are not shown here. In order that this works the important point is that the $c_i$ do not depend on the variable of integration. In general, the functions $f_i(x)$ are chosen to be derivatives or shifts in the parameter(s) of the integrand $f(x)$ if we are interested in differential equations or recurrences for the definite integral.

The main task for finding such relations of definite integrals of given functions $f_i(x)$ consists in finding suitable choices for the constants $c_i$ which allow a closed form of the antiderivative $g(x)$ to be computed. We will call this \emph{parametric integration} as it can be viewed as making suitable choices for the parameters $c_i$ occurring in the combined integrand $c_0f_0(x)+\dots+c_mf_m(x)$.

The approach above also addresses the issue of verifiability. When given such a linear relation of integrals
\[
 c_0\int_a^bf_0(x)\,dx+\dots+c_m\int_a^bf_m(x)\,dx=r
\]
the function $g(x)$ may act as a proof certificate of it as we just need to verify
\[
 c_0f_0(x)+\dots+c_mf_m(x)=g^\prime(x) \quad\quad\text{and}\quad\quad r=g(b)-g(a),
\]
where the left hand sides are directly read off from the integral relation we want to verify.

\subsection{Symbolic Integration}
\index{symbolic integration}

Algorithms to compute indefinite integrals of rational integrands are known for a long time already and many other integrals were computed analytically by hand as mentioned above. Especially in the last century algorithms have been developed capable of dealing with more general classes of integrands in a completely systematic way. In the following we want to give an overview of three different approaches that were taken. We also mention some relevant cornerstones but we do not include a fully comprehensive survey of the corresponding literature, many other contributions were made. Note that all of those approaches extend to definite integration in one way or the other.

The differential algebra approach represents functions as elements of differential fields and differential rings. These are algebraic structures not only capturing the arithmetic properties of functions but also their differential properties by including derivation as an additional unary operation. In general terms, starting with a prescribed differential field one is interested in indefinite integrals in the same field or in extensions of that field constructed in a certain way. Based on a book by Joseph~F.~Ritt \cite{RaabRef:Ritt} using differential fields Robert~H.~Risch gave a decision procedure \cite{RaabRef:Risch,RaabRef:BronsteinElem} for computing elementary integrals of elementary functions by closely investigating the structure of the derivatives of such functions. Since then this result has been extended in various directions. A parametric version was discussed in \cite{RaabRef:CMack}. Michael~F.~Singer et al.\ generalized this to a parametric algorithm computing elementary integrals over regular Liouvillian fields in the appendix of \cite{RaabRef:SingerEtAl} and Manuel Bronstein gave partial results for more general differential fields constructed by monomials \cite{RaabRef:BronsteinUni,RaabRef:Bronstein}. The author's thesis \cite{RaabRef:RaabPhD} can be seen as a continuation of this line of research. In \cite{RaabRef:Norman} Arthur~C.~Norman published a variant of Risch's algorithm avoiding its recursive structure, which therefore is sometimes also called the parallel Risch algorithm. The Risch-Norman algorithm can be used in even more general differential fields and has proven to be a rather powerful heuristic in practice, see \cite{RaabRef:Bronstein,RaabRef:BronsteinParallel,RaabRef:BoettnerPhD} and references therein. Most results mentioned so far restrict to the case where the generators of the differential fields are algebraically independent. The presence of algebraic relations causes new situations and requires more involved algebraic tools, see \cite{RaabRef:BronsteinElem,RaabRef:BronsteinTut,RaabRef:Kauers,RaabRef:BoettnerPhD} and references therein. Another type of generalization is to search also for certain types of non-elementary integrals over certain differential fields. Some results for this problem have been achieved in \cite{RaabRef:SingerEtAl}, see also \cite{RaabRef:Baddoura} and the references to the work of Cherry and Knowles in \cite{RaabRef:Bronstein}.

Indefinite integrals of products of special functions that satisfy homogeneous second-order differential equations were considered by Jean~C.~Piquette. His ansatz for the integral in terms of linear combinations of such products led to a differential system, which after uncoupling he solved by heuristic methods, see \cite{RaabRef:PiquetteBuren,RaabRef:Piquette} and references therein. The holonomic systems approach was initiated by Doron Zeilberger in \cite{RaabRef:Zeilberger} and puts this on more general and more algorithmic grounds. Functions are represented by the differential and difference operators that annihilate them. The notion of $D$-finite functions is closely related and refers to functions satisfying homogeneous linear differential equations with rational functions as coefficients. Hence, the derivatives of a $D$-finite function generate a finite-dimensional vector space over the rational functions. Fr\'ed\'eric Chyzak \cite{RaabRef:Chyzak} presented an efficient algorithm for computing indefinite integrals of such functions in the same vector space. The algorithm handles also parametric integration and summation and utilizes Ore algebras to represent the operators corresponding to functions. For extensions and improvements see \cite{RaabRef:Chyzak2,RaabRef:KoutschanPhD}.

The rule-based approach operates on the syntactic presentation of the integral by a table of transformation rules. This comes close to what is done when integrating by hand based on integral tables such as \cite{RaabRef:Integraltafel2,RaabRef:GR,RaabRef:PrudnikovEtAl}. Also most computer algebra systems make at least partial use of transformations and table look-up. These tables may contain rules for virtually any special function, which makes such algorithms easily extensible in principle. This approach is recently being investigated systematically by Albert~D.~Rich and David~J.~Jeffrey \cite{RaabRef:JeffreyRich1}, who point out several subtle issues related to efficiency.

\subsection{Risch's Algorithm}

When computing elementary integrals the paradigm followed by Risch's algorithm and many of its generalizations is that the computation proceeds recursively, focusing one by one on a particular function, which is involved in the integrands, at a time. For each of these functions the computation is organized in several main steps, where each step computes a part of the integral and subtracts its derivative from the integrand to obtain the remaining integrand to proceed with. The part of the integral that is computed in each step is chosen in such a way that the remaining integrand is simpler than the previous one in some suitable sense.

Before we discuss the computation for rather general types of integrands in a bit more detail, it will be instructive to consider the simplest case first, namely rational functions. The main steps of the full algorithm will be used to work out a closed form of the following integral of a rational function.
\[
 \int\frac{x^4+2x^3-x^2+3}{x^3+5x^2+8x+4}\,dx
\]
For rational functions three steps are relevant. First, we will apply Hermite reduction to reduce the task to an integrand that does not have poles of order greater than one. Then, we will compute the residues at the simple poles of the remaining integrand to obtain the logarithmic part of the integral. Finally, the remaining integrand will be a polynomial, which is easily integrated.

Let us start by outlining the main idea of Hermite reduction \cite{RaabRef:Hermite}, which repeats as needed what can be summarized as a suitably chosen additive splitting of the integrand followed by integration by parts of one of the two summands. Each time the order of some poles of the integrand is reduced. We will see later how such splittings are determined, for now we just emphasize that no partial fraction decomposition is required. In our example the denominator factors as $(x+1)(x+2)^2$. This means that we have to reduce the order of the pole at $x=-2$, which is achieved by the following splitting.
\begin{eqnarray*}
 \int\frac{x^4+2x^3-x^2+3}{(x+1)(x+2)^2}\,dx &=& \int\frac{1}{(x+2)^2}\,dx+\int\frac{x^3-x+1}{(x+1)(x+2)}\,dx\\
 &=& -\frac{1}{x+2}+\int\frac{x^3-x+1}{(x+1)(x+2)}\,dx.
\end{eqnarray*}

The remaining integrand has only simple poles, so we proceed by computing the residues at its poles, from which we obtain the logarithmic part of the integral. For an integrand $\frac{a(x)}{b(x)}$, where $a(x)$ and $b(x)$ are polynomials and $b(x)$ is squarefree, the residue at a root $x_0$ of $b(x)$ is given by $z_0:=\frac{a(x_0)}{b^\prime(x_0)}$ and we get a contribution $z_0\log(x-x_0)$ to the integral. Instead of determining the residue in dependence of the location of the pole, there are algorithms which first compute the set of values occurring as residues and then determine the appropriate logands for each residue. One such algorithm relying on resultants has its roots in the work of Rothstein and Trager, see \cite{RaabRef:Rothstein,RaabRef:Trager,RaabRef:LazardRioboo,RaabRef:Mulders}, and another algorithm using Gr\"{o}bner bases was proposed by Czichowski, see \cite{RaabRef:Czichowski}. The main idea to compute the residues directly, without computing the roots of $b(x)$, is to characterize them as those values $z_0$ such that the equations $a(x)-z_0{\cdot}b^\prime(x)=0$ and $b(x)=0$ are satisfied for some $x$ at the same time. So the Rothstein-Trager resultant $r(z)=\res_x(a(x)-z{\cdot}b^\prime(x),b(x))$ is a polynomial in the new variable $z$ having the residues of the integrand as its roots. For each root $z_0$ of $r(z)$ we need to compute $\gcd(a(x)-z_0{\cdot}b^\prime(x),b(x))$, which is the corresponding logand. A modern variant, the Lazard-Rioboo-Trager algorithm, computes the subresultant polynomial remainder sequence of $a(x)-z{\cdot}b^\prime(x)$ and $b(x)$, from which both $r(z)$ and the logands can be read off. Similarly, Czichowski's algorithm computes a Gr\"{o}bner basis of $\{a(x)-z{\cdot}b^\prime(x),b(x)\}$ w.r.t.\ $z<_{lex}x$, from which both the squarefree part of $r(z)$ and the logands can be read off. In our present example we determine the polynomials
\[
 r(z)=(z-1)(z-5) \quad\text{and}\quad s(z,x)=x+\frac{1}{4}z+\frac{1}{4},
\]
where the roots of $r(z)$ are the residues and $s(z,x)$ gives the corresponding logands, which give rise to the following logarithmic part of the integral.
\[
 \sum_{r(z)=0}z{\cdot}\log(s(z,x))=\log(x+1)+5\log(x+2)
\]
Therefore, subtracting its derivative from the integrand we obtain the polynomial
\[
 \frac{x^3-x+1}{(x+1)(x+2)} - \sum_{r(z)=0}z\frac{\frac{d}{dx}s(z,x)}{s(z,x)} = x-3
\]
as remaining integrand or, in other words,
\[
 \int\frac{x^3-x+1}{(x+1)(x+2)}\,dx = \log(x+1)+5\log(x+2)+\int{x-3}\,dx.
\]

The integral of a polynomial is determined via an appropriate ansatz, based on the fact that the derivative of a non-constant polynomial is a polynomial with degree exactly one less. The coefficients in the integral are then determined by equating the coefficients in the derivative of the ansatz to those in the integrand. In our case the ansatz for the integral is $a_2x^2+a_1x$ and comparing coefficients of the powers of $x$ in
\[
 \frac{d}{dx}\big(a_2x^2+a_1x\big) = x-3
\]
yields $2a_2=1$ and $a_1=-3$. Plugging the solution of these equations into the ansatz we obtain the integral
\[
 \int{x-3}\,dx = \frac{1}{2}x^2-3x.
\]

Altogether, we obtained the following closed form of the integral.
\[
 \int\frac{x^4+2x^3-x^2+3}{(x+1)(x+2)^2}\,dx = -\frac{1}{x+2}+\ln(x+1)+5\ln(x+2)+\frac{1}{2}x^2-3x
\]

\section{Algebraic Representation of Functions}
\label{RaabSec:AlgRep}

In differential algebra functions are represented as elements of differential fields and differential rings. These are algebraic structures not only capturing the arithmetic relations of functions but also their differential properties by including derivation as an additional operation. For more information on differential algebra, see \cite{RaabRef:Kaplansky} for example.

\begin{definition}
 Let $F$ be a field and let $D : F \to F$ be a unary operation on it, which is additive and satisfies the product rule, i.e.,
 \[
  D(f+g)=Df+Dg \quad\quad\mbox{and}\quad\quad D(fg)=fDg+(Df)g.
 \]
 Then $D$ is called a \emph{derivation} on $F$ and $(F,D)$ is called a \index{differential field}\emph{differential field}. The set of \emph{constants} is denoted by $\const_D(F):=\set{f \in F}{Df=0}$.
\end{definition}

It follows from the definition that the set $\const_D(F)$ is closed under the basic arithmetic operations and hence is a field. Note that while for $f,g \in F$ sums $f+g$, products $fg$, and derivatives $Df$ by definition are in $F$ again, powers $f^g$, compositions $f \circ g$ and antiderivatives $\int f$ need not be in $F$ in general. The same statements apply for differential rings where every occurrence of the word field is to be replaced by the word ring.\par
The basic example for a differential field is the field of rational functions $(F,D)=(C(x),\frac{d}{dx})$, where $Dx=1$ and $\const_D(F)=C$. Note that $C$, the field of constants, may not only consist of numbers, but it may also contain elements depending on variables other than $x$. For example, $(F,D)=(\mathbb{R}(n,x,x^n,\ln(x)),\frac{d}{dx})$, where the notation is meant to imply $Dn=0$, $Dx=1$, $Dx^n=\frac{n}{x}x^n$, and $D\ln(x)=\frac{1}{x}$, is a differential field with $\const_D(F)=\mathbb{R}(n)$. In principle the definition of a differential field does not require the existence of an element $x \in F$ with $Dx=1$. For example, $(\mathbb{Q}(e^x),\frac{d}{dx})$ is a differential field since the derivative of any rational expression in $e^x$ is again a rational expression in $e^x$. In practice, however, such cases are not very important.\par
In general, we will consider finitely generated differential fields of the form $(F,D)=(C(t_1,\dots,t_n),D)$, where $C=\const_D(F)$ and $t_1,\dots,t_n$ represent some functions. Algebraically, an element $f \in F$ is a rational expression in $t_1,\dots,t_n$ with coefficients in $C$ and, resembling the chain rule, the derivative can be expressed as
\[
 Df = \frac{\partial f}{\partial t_1}\cdot Dt_1 + \dots + \frac{\partial f}{\partial t_n}\cdot Dt_n.
\]

When given some differential field $(K,D)$, by adjoining additional elements $t_1,\dots,t_n$ under the condition that we can extend $D$ to a derivation on $K(t_1,\dots,t_n)$ we obtain a differential field extension $(K(t_1,\dots,t_n),D)$, i.e., a differential field containing $(K,D)$ as a differential subfield. The following theorem makes the choice explicit which we have when extending the derivation from $(K,D)$ to a differential field extension of the form $(K(t),D)$.

\begin{theorem}(\cite[Theorems~3.2.2, 3.2.3]{RaabRef:Bronstein})\index{Liouville's Theorem}
 Let $(K,D)$ be a differential field and let $K(t)$ be the field generated by a new element $t$.
 \begin{enumerate}
  \item If $t$ is algebraic over $K$, then $D$ can be uniquely extended to a derivation on $K(t)$.
  \item If $t$ is transcendental over $K$, then, for any $w \in K(t)$, $D$ can be uniquely extended to a derivation on $K(t)$ such that $Dt=w$.
 \end{enumerate}
\end{theorem}

In our presentation we will focus on transcendental extensions, in which case the notion of a (differential) monomial, introduced by Bronstein \cite{RaabRef:BronsteinUni}, is very important for practical algorithms.

\begin{definition}\label{RaabDef:Monomial}
 Let $(F,D)$ be a differential field, $(K,D)$ a differential subfield, and $t \in F$. Then $t$ is called a \index{monomial extension}\emph{monomial} over $(K,D)$ if
 \begin{enumerate}
  \item $t$ is transcendental over $K$ and
  \item $Dt \in K[t]$.
 \end{enumerate}
 If $\deg_t(Dt)\ge2$, we call $t$ \emph{nonlinear}.
\end{definition}

There are many similarities between rational functions $(C(x),\frac{d}{dx})$ and monomial extensions $(K(t),D)$, but there are, of course, some important differences as well. The derivative of polynomials $p \in C[x]$ is a polynomial again, likewise any polynomial $p \in K[t]$ has its derivative $Dp$ in $K[t]$. However, unlike the degree of the derivative of $p \in C[x]$, the degree need not drop when applying $D$ to a  $p \in K[t]$, it may stay the same or even increase depending on the degree in $t$ of $Dt$. An irreducible polynomial $p \in C[x]$ never divides its derivative, this need not be true for polynomials $p \in K[t]$. More generally, a squarefree polynomial $p \in K[t]$ need not be coprime with $Dp$, while it always is if $p \in C[x]$.
\begin{definition}\label{RaabDef:NormalSpecial}
Let $(K,D)$ be a differential field and let $t$ be a monomial over $(K,D)$. We call a polynomial $p \in K[t]$ \emph{normal}, if $p$ and $Dp$ are coprime, or \emph{special}, if $p$ divides $Dp$.
\end{definition}
A squarefree polynomial $p \in K[t]$ is normal if and only if it does not contain a factor of degree at least $1$ which is special.\par
These properties of the derivation on $K[t]$ deserve to be exemplified, for which we consider $(K,D)=(C(x),\frac{d}{dx})$. The transcendental functions $\ln(x)$, $\exp(x)$, and $\tan(x)$ satisfy $\frac{d}{dx}\ln(x)=\frac{1}{x}$, $\frac{d}{dx}\exp(x)=\exp(x)$, and $\frac{d}{dx}\tan(x)=\tan(x)^2+1$, respectively. Hence they can be represented by monomials over $(K,D)$. Representing the logarithm by a monomial $t$ with $Dt=\frac{1}{x} \in K$ it can be proven that the degree of a polynomial $p \in K[t]$ is always at least as large as that of its derivative $Dp$. The degrees are unequal if and only if the leading coefficient of $p$ is in $C$, which is true for $p=t^2+xt$ with $Dp=\frac{x+2}{x}t+1$, for example, but not for $p=xt^2-\frac{2x^2-x}{x+1}t$ with $Dp=t^2+\frac{3}{(x+1)^2}t-\frac{2x-1}{x+1}$. In addition, every squarefree polynomial is indeed coprime with its derivative. For a monomial $t$ with $Dt=t$ we always have that $p \in K[t]$ and $Dp$ have the same degree if $p$ is not constant. There are polynomials which divide their derivative, and all of them are of the form $p=at^n$ where $a \in K$ and $n \in \mathbb{N}$. Finally, a monomial with $Dt=t^2+1$ has the property that the degree of $Dp$ is strictly greater than that of $p \in K[t]$ as long as the degree of $p$ is at least one, e.g., $p=t(t^2+1)$ has derivative $Dp=(3t^2+1)(t^2+1)$, and a squarefree polynomial is normal if and only if it is coprime with $t^2+1$.

Furthermore, in monomial extensions $(K(t),D)$ we will rely on the \emph{canonical representation}
\[
 f=p+\frac{a_s}{b_s}+\frac{a_n}{b_n}
\]
of elements $f \in K(t)$, where $p,a_s,a_n,b_s,b_n \in K[t]$ with $\deg_t(a_s)<\deg_t(b_s)$ and $\deg_t(a_n)<\deg_t(b_n)$ are such that $b_s$ is special and every irreducible factor of $b_n$ is normal, cf.~\cite[ p.103]{RaabRef:Bronstein}

\subsection{Relevant Classes of Functions}

Apart from rational functions $(C(x),\frac{d}{dx})$ and algebraic functions $(\overline{C(x)},\frac{d}{dx})$, elementary functions are a very basic class of functions as well and were among the first to be considered algorithmically. The \index{elementary functions}\emph{elementary functions} are those which can be constructed from rational functions by the following operations in addition to the basic arithmetic operations: taking the logarithm, applying the exponential function, and solving algebraic equations with elementary functions as coefficients. Elementary functions include rational and algebraic functions, logarithms, $c^x$ and $x^c$, trigonometric functions and their inverses, as well as hyperbolic functions and their inverses. Recall that trigonometric and hyperbolic functions can be expressed in terms of exponentials and their inverses can be expressed in terms of logarithms of algebraic functions. Note that compositions $f(g(x))$ and powers $f(x)^{g(x)}$ of elementary functions are elementary functions again. When representing elementary functions in differential fields we make use of the following relations.
\begin{eqnarray}
 \frac{d}{dx}\ln(a(x))&=&\frac{a^\prime(x)}{a(x)}\label{RaabEq:Logarithm}\\
 \frac{d}{dx}\exp(a(x))&=&a^\prime(x)\exp(a(x))\label{RaabEq:Exponential}
\end{eqnarray}

\begin{definition}
 Let $(K,D)$ be a differential field and let $t$ be a monomial over $(K,D)$. Then we call $t$ an \emph{elementary monomial} over $(K,D)$ if it is either
 \begin{enumerate}
  \item a \emph{logarithm} over $(K,D)$, i.e., there exists $a \in K$ such that $Dt=\frac{Da}{a}$, or
  \item an \emph{exponential} over $(K,D)$, i.e., there exists $a \in K$ such that $\frac{Dt}{t}=Da$.
 \end{enumerate}
 Let $(F,D)=(K(t_1,\dots,t_n),D)$ be a differential field extension of $(K,D)$. Then $(F,D)$ is called \emph{elementary extension} of $(K,D)$, if each $t_i$ is either algebraic or an elementary monomial over $(K(t_1,\dots,t_{i-1}),D)$.
\end{definition}

An elementary function is a function representable as an element of some elementary extension of $(C(x),\frac{d}{dx})$. Note that an elementary extension of some differential field $(K,D)$ does not only contain elementary functions unless $K$ does.

The notion of elementary functions is generalized naturally to give \index{Liouvillian functions}\emph{Liouvillian functions} by considering differential equations of the form
\begin{eqnarray}
 \frac{d}{dx}y(x)&=&a(x)\label{RaabEq:Primitve}\\
 \frac{d}{dx}y(x)&=&a(x)y(x)\label{RaabEq:Hyperexponential}
\end{eqnarray}
instead of their special cases for logarithms and exponentials above. In other words, Liouvillian functions are the functions obtained from rational functions by the basic arithmetic operations, by taking primitive functions $\int a(x)\,dx$, by taking hyperexponential functions $e^{\int a(x)\,dx}$, and by solving algebraic equations with Liouvillian functions as coefficients. Again, the composition of Liouvillian functions as well as powers $f(x)^{g(x)}$ of Liouvillian functions are Liouvillian. Several special functions can be found in the class of Liouvillian functions, e.g., logarithmic and exponential integrals, error functions, Fresnel integrals, incomplete Beta and $\Gamma$ functions, polylogarithms, harmonic polylogarithms \cite{RaabRef:RemiddiVermaseren}, and hyperlogarithms \cite{RaabRef:Brown}.

\begin{definition}
 Let $(K,D)$ be a differential field and let $t$ be a monomial over $(K,D)$. Then we call $t$ a \emph{Liouvillian monomial} over $(K,D)$ if it is either
 \begin{enumerate}
  \item \emph{primitive} over $(K,D)$, i.e., there exists $a \in K$ such that $Dt=a$, or
  \item \emph{hyperexponential} over $(K,D)$, i.e., there exists $a \in K$ such that $\frac{Dt}{t}=a$.
 \end{enumerate}
 Let $(F,D)=(K(t_1,\dots,t_n),D)$ be a differential field extension of $(K,D)$. Then $(F,D)$ is called \emph{Liouvillian extension} of $(K,D)$, if each $t_i$ is either algebraic or a Liouvillian monomial over $(K(t_1,\dots,t_{i-1}),D)$.
\end{definition}

A Liouvillian function is a function representable as an element of some Liouvillian extension of $(C(x),\frac{d}{dx})$. Note that there are a few equivalent definitions of the class of Liouvillian functions. For instance, we need not start the construction from the rational functions but it suffices to start from the set of constants because the rational functions are obtained by the basic arithmetic operations from constants and the identity function, which in turn is a primitive function of the constant $1$. Similarly, we may also choose to keep the operation of applying the exponential function instead of replacing it by taking hyperexponential functions as the latter operation can obviously be decomposed into applying the exponential function to a primitive function. Alternatively, we may also summarize taking primitive and hyperexponential functions into taking solutions of linear first-order differential equations. More precisely, the class of Liouvillian functions may also be constructed from the set of constants by the basic arithmetic operations and taking particular solutions of
\begin{equation}\label{RaabEq:Liouvillian}
 y^\prime(x)=a(x)y(x)+b(x)
\end{equation}
and of algebraic equations with Liouvillian coefficients each. Note that the solutions of \eqref{RaabEq:Liouvillian} may be expressed in terms of primitives and (hyper)exponentials by $y(x)=e^{\int a(x)\,dx}\int\frac{b(x)}{e^{\int a(x)\,dx}}\,dx$.

For the sake of completeness we also give the definition of hyperexponential and d'Alembertian functions \cite{RaabRef:Abramov93,RaabRef:AbramovPetkovsek}, although they are not so relevant in our considerations. They are continuous analogues of hypergeometric and d'Alembertian sequences, respectively. An algorithm for integration of hyperexponential functions is given in \cite{RaabRef:AlmkvistZeilberger}.
\begin{definition}
Let $(F,D)$ be a differential field, $(K,D)$ a differential subfield, and $t \in F$. Then $t$ is called
 \begin{enumerate}
  \item \emph{hyperexponential} over $(K,D)$ if $\frac{Dt}{t} \in K$, or
  \item \emph{d'Alembertian} over $(K,D)$ if there exist $n \in \mathbb{N}$ and $r_1,\dots,r_n \in K$ such that $t$ is a solution of the homogeneous linear differential equation obtained from the composition of differential operators $D-r_i$, i.e., $(D-r_n)\dots(D-r_1)t=0$.
 \end{enumerate}
\end{definition}

The \index{hyperexponential functions}\emph{hyperexponential functions} are functions $h(x)$ being hyperexponential over $(C(x),\frac{d}{dx})$, i.e., their logarithmic derivative $\frac{h^\prime(x)}{h(x)}$ is a rational function. Typical examples of hyperexponential functions are $c^{f(x)}$ and $f(x)^c$, where $f(x)$ is a rational function. Note that the product and the quotient of hyperexponential functions are hyperexponential again, but the sum of hyperexponential functions is not hyperexponential in general. So, in contrast to the classes of elementary and Liouvillian functions, the class of hyperexponential functions is not closed under the basic arithmetic operations.

Similarly, \index{d'Alembertian functions}\emph{d'Alembertian functions} are the functions $h(x)$ that are d'Alembertian over $(C(x),\frac{d}{dx})$. The class of d'Alembertian functions is not closed under the basic arithmetic operations either, as the sum and the product of d'Alembertian functions are d'Alembertian again, but the quotient of d'Alembertian functions is not d'Alembertian in general. Most of the special functions listed above as being Liouvillian functions are in fact even d'Alembertian functions: exponential integrals, error functions, Fresnel integrals, incomplete Beta and $\Gamma$ functions, polylogarithms, harmonic polylogarithms, and hyperlogarithms. Note that hyperexponential functions are d'Alembertian as well, and d'Alembertian functions are Liouvillian. An equivalent characterization of d'Alembertian functions is that they can be written as iterated integrals over hyperexponential functions
\[
 h_1(x)\int h_2(x)\int\dots\int h_n(x)\,dx\dots dx.
\]
The relation to the previous definition is that the product $h_1(x)\dots h_i(x)$ is a solution of $y^\prime(x)-r_i(x)y(x)=0$.

\subsection{Liouville's Theorem}

Liouville \cite{RaabRef:Liouville1,RaabRef:Liouville1a,RaabRef:Liouville2} was the first to prove an observation on the structure of elementary integrals. In the language of differential fields it can be stated as follows.

\begin{theorem}(Liouville's Theorem \cite[Thm~5.5.3]{RaabRef:Bronstein})
 Let $(F,D)$ be a differential field and $C:=\const(F)$. If $f \in F$ has an elementary integral over $(F,D)$, then there are $v \in F$, $c_1,\dots,c_n \in \overline{C}$, and $u_1,\dots,u_n \in F(c_1,\dots,c_n)^*$ such that
 \begin{equation}
  f=Dv+\sum_{i=1}^nc_i\frac{Du_i}{u_i}.
 \end{equation}
\end{theorem}

In view of this theorem we always can express an elementary integral $\int f$ as the sum of two parts: a $v \in F$, which then is called the \emph{rational part}, and a sum of logarithms $\sum c_i\log(u_i)$, which is called the \emph{logarithmic part} of the integral. This theorem and its refinements \cite{RaabRef:Bronstein,RaabRef:RaabPhD} which consider a special structure of the integrand are the main theoretical foundation for algorithms computing elementary integrals. There are even generalizations of Liouville's theorem dealing also with non-elementary integrals, e.g.\ \cite{RaabRef:SingerEtAl,RaabRef:Baddoura}.

\section{Risch's Algorithm in Monomial Extensions}
\label{RaabSec:Alg}

As already explained earlier, we are interested in parametric integration. In terms of differential fields this problem can be formulated as follows.

\begin{problem}[parametric elementary integration]\label{RaabProb:elemint}\index{parametric integration}
Given: a differential field $(F,D)$ and $f_0,\dots,f_m \in F$.\\[2mm]
Find: a $C$-vector space basis $\mathbf{c}_1,\dots,\mathbf{c}_n \in C^{m+1}$, where $C:=\const(F)$, of all coefficient vectors $(c_0,\dots,c_m) \in C^{m+1}$ such that $c_0f_0+\dots+c_mf_m \in F$ has an elementary integral over $(F,D)$ and compute corresponding integrals $g_1,\dots,g_n$ from some elementary extensions of $(F,D)$.
\end{problem}

We consider this problem over towers of \index{monomial extension}monomial extensions, i.e., $(F,D)=(C(t_1,\dots,t_n),D)$ where each $t_i$ is a monomial over $(C(t_1,\dots,t_{i-1}),D)$ subject to some technical conditions. For details see \cite{RaabRef:Bronstein,RaabRef:RaabPhD}. A big part of the common special functions can be represented in such differential fields. In addition to Liouvillian functions, most importantly functions satisfying (possibly inhomogeneous) linear second-order differential equations can be fit into this framework. Concrete examples include orthogonal polynomials, associated Legendre functions, Bessel functions, Airy functions, complete elliptic integrals, Whittaker functions, Mathieu functions, hypergeometric functions, Heun functions, Struve functions, Anger functions, Weber functions, Lommel functions, Scorer functions, etc. How this can be done is explained in \cite{RaabRef:RaabPhD}.

As mentioned above Risch's algorithm proceeds recursively, thereby exploiting the structure of the underlying differential field that is used to model the functions occurring. The focus of the computation always is on the topmost generator of the differential field and everything else is regarded as part of the coefficients. In essence, the steps dealing with expressions from $C(x)$ outlined above are generalized to work with expressions from $K(t)$ where some monomial $t$, cf.\ Defintion~\ref{RaabDef:Monomial}, takes the role of $x$ and coefficients appearing in rational or polynomial expressions in $t$ do not necessarily have zero derivative. Moreover, we do not consider the poles of the integrand by interpreting it as a function of $x$, we will work on a syntactic level instead by considering the factors of the denominator in the representation of the integrand in terms of $t$. The algorithms outlined above carry over as long as they are applied to the normal part of the denominator only. If present, the special part of the denominator needs to be treated differently, which is done similarly to integrating the polynomial part.

Along with the main ideas of the algorithm in monomial extensions we present a specific example to illustrate how the integrand is processed. For the explicit computation we consider the integral
\[
 \int\frac{x^2 e^{5x} - 2 x e^{4x} + (2 x^3 + 5 x + 1) e^{3x} - (6 x^3 + x + 1) e^x + 4 x^3}{x^2 e^{2x} (e^x - 1)^2}\,dx.
\]
The integrand can be represented in the differential field $(C(x,t),D)$ with $Dx=1$ and $Dt=t$ as
\[
 \frac{x^2 t^5 - 2 x t^4 + (2 x^3 + 5 x + 1) t^3 - (6 x^3 + x + 1) t + 4 x^3}{x^2 t^2 (t - 1)^2}.
\]

In the general setting Hermite reduction requires some preprocessing, since it only deals with terms for which all irreducible factors of the denominator are normal. To this end, we compute the canonical representation mentioned earlier. We ignore any terms with special polynomials in the denominator for the moment.

In our example we have that the polynomial $t$ is special and the polynomial $t-1$ is normal. So the canonical representation is given by
\[
 t+\frac{2x-2}{x}+\frac{\frac{2x^3-x-1}{x^2}t+4x}{t^2}+\frac{\frac{3x^2+2x+2}{x^2}t-\frac{2x^2+2}{x^2}}{(t-1)^2},
\]
where the last fraction is the one we will focus on now.

Hermite reduction repeatedly splits the integrand and applies integration by parts to one of the two summands each time. More precisely, if the integrand is of the form $\frac{a}{uv^{m+1}}$, where $a,u,v \in K[t]$ are pairwise relatively prime polynomials with $v$ being normal and $m \in \{2,3,\dots\}$, then there are unique polynomials $r,s \in K[t]$ such that $\deg_t(r)<\deg_t(v)$ and
\[
 a = (1-m)ruDv+sv.
\]
Such polynomials can be readily computed by the extended euclidean algorithm, for instance. With this splitting of the numerator we have
\begin{equation}\label{RaabEq:HermiteRational}
 \int\frac{(1-m)ruDv+sv}{uv^m} = \frac{r}{v^{m-1}} + \int\frac{s-uDr}{uv^{m-1}},
\end{equation}
where the power of $v$ in the denominator of the remaining integrand has dropped by (at least) one. Note that the polynomial $v$ is merely required to be normal, so all normal irreducible factors in the denominator of the integrand occurring with power $m$ can be treated at once. 

Hermite reduction repeats the above step until an integrand with a normal denominator is obtained. Starting from an integrand $\frac{a}{b}$ with $a,b \in K[t]$ and every irreducible factor of $b$ being normal, we first compute a squarefree factorization of the denominator $b=b_1b_2^2\dots b_n^n$ and then after at most $n-1$ reduction steps going from $m=n$ down to $m=2$, reducing the highest-order poles in each step, we arrive at an integrand with a normal denominator.

There is also a variant of Hermite reduction where at each reduction step the order of all poles of order greater than one is reduced, instead of the highest-order poles only. This has the additional advantage that no squarefree factorization needs to be computed at the beginning.

In our example the denominator $(t-1)^2$ is already given in factored form. This means that we have $m=2$, $u=1$, and $v=t-1$. With these values we need to find the polynomials $r,s \in C(x)[t]$ satisfying
\[
 \frac{3x^2+2x+2}{x^2}t-\frac{2x^2+2}{x^2} = r\cdot(-t)+s\cdot(t-1)
\]
and $\deg_t(r)<1$. We compute $r(x)=-\frac{x+2}{x}$ and $s=\frac{2x^2+2}{x^2}$, so by \eqref{RaabEq:HermiteRational} we obtain
\[
 \int\frac{\frac{3x^2+2x+2}{x^2}t-\frac{2x^2+2}{x^2}}{(t-1)^2} = -\frac{x+2}{x(t-1)}+\int\frac{2}{t-1}.
\]

The remaining integrand has a normal denominator and we still focus on the part of the integrand which has normal irreducible factors in its denominator only, which just occur with multiplicity one now. For such integrands the notion of a residue can be defined appropriately in monomial extensions, which we do not detail here. We proceed by computing the logarithmic part of the integral, which will be of the form
\[
 \sum_i\sum_{r_i(z)=0}z{\cdot}\log(s_i(z,t))
\]
with $r_i \in C[z]$ squarefree and $s_i \in K[z,t]$. This means that the residues are the roots of the polynomials $r_i$ and the polynomials $s_i$ give the corresponding logands. In general it may happen that the residue is not a constant, i.e., potentially we have $r_i \in K[z]$ only. If this happens, it can be shown that the integral is not elementary over $(K(t),D)$. This gives a necessary condition on the coefficients of the linear combination of several integrands in the parametric integration problem. An algorithm to ensure that we will consider only linear combinations which actually have $r_i \in C[z]$ can be found in \cite{RaabRef:RaabPhD}, a different algorithm was already used in \cite{RaabRef:SingerEtAl}. Once this is done we compute the corresponding polynomials $r_i$ and $s_i$ via generalizations of the algorithms mentioned earlier that originally were designed for rational functions, see \cite{RaabRef:Bronstein,RaabRef:Raab1}. Note that subtracting the derivative
\[
 \sum_i\sum_{r_i(z)=0}z{\cdot}\frac{D(s_i(z,t))}{s_i(z,t)}
\]
of the logarithmic part of the integral from the integrand may also change the polynomial part of the integrand in the general case, in particular this happens if $t$ is nonlinear. 

In our case we simply have one polynomial $r_1=z-2$ and $s_1=t-1$ each, which give rise to the logarithmic part
\[
 2\log(t-1)
\]
Subtracting its derivative $D(2\log(t-1))=2+\frac{2}{t-1}$ from the integrand we obtain
\[
 t+\frac{2x-2}{x}+\frac{\frac{2x^3-x-1}{x^2}t+4x}{t^2}+\frac{2}{t-1} - \left(2+\frac{2}{t-1}\right) = t-\frac{2}{x}+\frac{\frac{2x^3-x-1}{x^2}t+4x}{t^2}.
\]

At this point the remaining integrands are such that their denominator is special. Depending on the specific properties of $t$ this condition admits only a very restricted form of the denominator and in many cases even implies that the denominator is in $K$. The aim is to reduce the integrands to lie in $K$. In short, the idea how to proceed is to make an appropriate ansatz for part of the integral based on the partial fraction decomposition of the integrands. Comparing coefficients then leads to differential equations with coefficients in $K$, for which solutions have to be found in $K$. While setting up the ansatz and solving for the coefficients was the easiest part in the integration of rational functions, it is the most difficult part in the general setting and algorithms exist only for certain types of monomials $t$ and underlying differential fields $(K,D)$. Under certain technical assumptions on $t$ the following ansatz for the part of the integrands having special denominators can be justified. The integrand on the left hand side has only irreducible polynomials $p_j \in K[t]$ in its denominator which are special and it is given by its partial fraction decomposition.
\[
 \sum_{j=1}^n\sum_{k=1}^{l_j}\frac{f_{j,k}}{p_j^k}=D\left(\sum_{j=1}^n\sum_{k=1}^{l_j}\frac{g_{j,k}}{p_j^k}\right)
\]
After rewriting the right hand side in its partial fraction decomposition we can compare coefficients in order to obtain differential equations for $g_{j,k} \in K[t]$. Note that the derivative
\[
 D\frac{g_{j,k}}{p_j^k}=\frac{Dg_{j,k}-k\frac{Dp_j}{p_j}g_{j,k}}{p_j^k}
\]
again has the same power $p_j^k$ in the denominator since $\frac{Dp_j}{p_j} \in K[t]$ for special polynomials. Roughly speaking, upon comparing coefficients of $p_j^{-k}$ we obtain differential equations relating each $g_{j,k}$ to $f_{j,k}$. This leads to the problem of finding solutions of certain type to differential equations, which may or may not exist. If no solution of the correct type exists, then it can be shown that the integral is not elementary over $(K(t),D)$. This again restricts the possible linear combinations in the parametric integration problem. There is a lot more to this, but we do not go into detail here. Instead we refer to \cite{RaabRef:Bronstein} where relevant results are given. Not all cases can be dealt with algorithmically so far, this depends on the structure of $(K,D)$ as well as on $t$. The main difficulty lies in the algorithmic solution of the differential equations arising, for which we also refer to \cite{RaabRef:Abramov,RaabRef:Singer91,RaabRef:BronsteinODE} for example. This can be skipped if $t$ is such that $K[t]$ does not contain any special irreducible polynomial. In practice this is often the case, the most notable exception are hyperexponential monomials $t$.\par
The above ansatz deals with the remaining denominators in the integrands. Similarly, for the remaining polynomial parts we can set up an ansatz of the form
\[
 \sum_{j=1}^n f_jt^j= D\left(\sum_{j=1}^{n+1-d}g_jt^j\right)
\]
where $d:=\deg_t(Dt)$ and $g_j \in K$. After expanding the right hand side in powers of $t$, we compare coefficients of $t^{\max(d,1)},\dots,t^{n+\max(1-d,0)}$. The degree $d$ of $Dt$ determines the main features of the action of the derivation on polynomials from $K[t]$. If $t$ is nonlinear, i.e., $d\ge2$, then we can directly solve for $g_j$ one by one. Otherwise, this leads to differential equations for $g_j$, which again impose restrictions on the possible linear combinations of integrands. As above, depending on the structure of $(K,D)$ as well as on $t$ the algorithms for computing solutions to these differential equations given in \cite{RaabRef:Abramov,RaabRef:Singer91,RaabRef:BronsteinODE,RaabRef:Bronstein} apply. There are large classes relevant in practice, which can be solved completely algorithmically. Remaining integrands are polynomials in $K[t]$ of degree less than $\max(d,1)$, which can be reduced further to integrands in $K$ under certain assumptions on $t$.

Our running example is such that complete algorithms exist. The fractional part has partial fraction decomposition
\[
 \frac{\frac{2x^3-x-1}{x^2}t+4x}{t^2}=\frac{2x^3-x-1}{x^2t}+\frac{4x}{t^2}
\]
with respect to $t$. The ansatz $\frac{g_1}{t}+\frac{g_2}{t^2}$ has derivative $\frac{Dg_1-g_1}{t}+\frac{Dg_2-2g_2}{t^2}$ and hence leads to the differential equations
\begin{eqnarray*}
 Dg_1-g_1&=&\frac{2x^3-x-1}{x^2}\\
 Dg_2-2g_2&=&4x
\end{eqnarray*}
with solutions $g_1=-\frac{2x^2+2x-1}{x} \in C(x)$ and $g_2=-2x-1 \in C(x)$. The polynomial part is just $t$, for which the ansatz $g_1t$ for the integral trivially leads to $g_1=1$. Altogether, we have the remaining integrand
\[
 t-\frac{2}{x}+\frac{\frac{2x^3-x-1}{x^2}t+4x}{t^2}-D\left(t-\frac{2x^2+2x-1}{xt}-\frac{2x+1}{t^2}\right)=\frac{2}{x} \in C(x).
\]

Now we reduced to integrands in $K$, still we want to find integrals which are elementary over $(K(t),D)$. If $t$ is elementary over $(K,D)$, then this obviously is equivalent to finding integrals elementary over $(K,D)$. In order to apply our algorithm recursively we have to reduce this to a problem of finding elementary integrals over $(K,D)$ also in the case where $t$ is non-elementary over $(K,D)$. Various refinements of Liouville's theorem are needed to solve this issue. For details we refer to \cite{RaabRef:RaabPhD}, we just mention that this may lead to an increase in the number of integrands we have to consider in the recursive application of the algorithm.

In case of our example $t$ is elementary over $(K,D)=(C(x),\frac{d}{dx})$, so we just need to apply the algorithm recursively to the remaining integrand $-\frac{2}{x}$. This yields $-2\log(x)$ as elementary integral over $(C(x),\frac{d}{dx})$. Now, collecting all the parts of the integral we computed, we obtain the following closed form
\[
 -2\log(x)+t-\frac{2x^2+2x-1}{xt}-\frac{2x+1}{t^2}+2\log(t-1)-\frac{x+2}{x(t-1)}.
\]
In other words we computed\\[\medskipamount]
$\displaystyle
 \int\frac{x^2 e^{5x} - 2 x e^{4x} + (2 x^3 + 5 x + 1) e^{3x} - (6 x^3 + x + 1) e^x + 4 x^3}{x^2 e^{2x} (e^x - 1)^2}\,dx =
$\\[\smallskipamount]\mbox{}\hfill$\displaystyle
 2\ln\left(\frac{e^x-1}{x}\right)+\frac{xe^{4x}-xe^{3x}-(2x^2+3x+1)e^{2x}+(x-1)e^x+x}{xe^{2x}(e^x-1)}.
$

\subsection{Non-monomial extensions}

To a certain extent the algorithm can also be applied even in situations where the differential field does not meet all the requirements. Depending on which properties are violated the computation still may make sense, for instance if some algebraic relations among the generators of the differential field exist. Then it is just not guaranteed to find all possible solutions. Recently this heuristic has proven to be quite effective in the computation of massive Feynman diagrams at 3-loops \cite{RaabRef:massive3loop} where new iterated integrals involving square-root terms emerged \cite{RaabRef:BinoialSums}.

\begin{acknowledgement}
The author was supported by the Austrian Science Fund (FWF), grant no.\ W1214-N15 project DK6, by the strategic program {\lq\lq}Innovatives O\"{O} 2010 plus{\rq\rq} of the Upper Austrian Government, by DFG Sonderforschungsbereich Transregio 9 {\lq\lq}Computergest\"{u}tzte Theoretische Teilchenphysik{\rq\rq}, and by the Research Executive Agency (REA) of the European Union under the Grant Agreement number PITN-GA-2010-264564 (LHCPhenoNet).
\end{acknowledgement}

\end{document}